\documentclass[twocolumn,prl,aps,superscriptaddress,showpacs]{revtex4}

\usepackage{bm}\usepackage{graphicx}

\begin{document}

\title{Further analysis of the quantum critical point of Ce$_{1-x}$La$_{x}$Ru$_{2}$Si$_{2}$}

\author{St\'ephane Raymond}
\affiliation{CEA-Grenoble, DRFMC / SPSMS / MDN, 38054 Grenoble Cedex, France}
\author{William Knafo\footnote{Present Address : FZK, Instit\"ut f\"ur Festk\"orperphysik, D-76021, Karlsruhe, Germany}}
\affiliation{CEA-Grenoble, DRFMC / SPSMS / MDN, 38054 Grenoble Cedex, France}
\author{Jacques Flouquet}
\affiliation{CEA-Grenoble, DRFMC / SPSMS / MDN, 38054 Grenoble Cedex, France}
\author{Pascal Lejay}
\affiliation{CRTBT, CNRS, 38042 Grenoble Cedex, France}
\date{\today}

\begin{abstract}
New data on the spin dynamics and the magnetic order of Ce$_{1-x}$La$_{x}$Ru$_{2}$Si$_{2}$ are presented. The importance of the Kondo effect at the quantum critical point of this system is emphasized from the behaviour of the relaxation rate at high temperature and from the variation of the ordered moment with respect to the one of the N\'eel temperature for various $x$. 

\end{abstract}

\pacs{71.27.+a, 75.40.Gb, 78.70.Nx}
\maketitle

\section{INTRODUCTION}

Recent leading topics in heavy fermion (HF) compounds physics are unconventional superconductivity and marginal Fermi liquid behaviour. Since these phenomena often occur in the vicinity of a magnetic instability, the concept of quantum phase transition has attracted a lot of attention both on experimental\cite{Stewart,Lohneysen} and theoretical\cite{Votja,Sachdev,Continentino} sides. Such a transition between a magnetically ordered state and a paramagnetic state occurs at $T$=0 K for a critical value, $r_{c}$, of a non thermal tuning parameter $r$ ($r$ being experimentally controlled by pressure, alloying or magnetic field). The $T$=0 K, $r$=$r_{c}$ point of the phase diagram is called the quantum critical point (QCP).
Inelastic neutron scattering (INS) is a key probe to study spin fluctuations at a QCP in order to know to which extend the characteristics of the dynamical spin susceptibility can explain the anomalous bulk properties of these materials. Peculiar behavior is expected because statics and dynamics are inextricably mixed at $T$=0 K. As a consequence, the effective dimension of the system $d_{eff}$ equals $d+z$, where $d$ is the space dimension and $z$ is the dynamical exponent \cite{Hertz,Millis}. This implies that a quantum critical system is in better position to be above the upper critical dimension than a classical one. Thus a mean field behavior is therefore expected at the QCP. While many INS studies have been performed near the QCP, only a few works were extensively performed on single crystal samples where the full wavewector ($\mathbf{q}$) dependence can be investigated. Such studies concern CeCu$_{6}$ doped with Au (Karlsruhe group), CeRu$_{2}$Si$_{2}$ doped with La (Grenoble group) and doped with Rh (Tokyo-Osaka group).

For the system Ce$_{1-x}$La$_{x}$Ru$_{2}$Si$_{2}$, large single crystals of good quality are available for several concentrations $x$ \cite{Lejay}. The three dimensional Ising compound CeRu$_{2}$Si$_{2}$ \cite{Flouquet} is an archetypal HF system with a Pauli paramagnetic ground state. Strong correlations between lattice, electronic and magnetic properties are evidenced under magnetic field where pseudo-metamagnetic transition occurs at $H_{m}$=7.8 T and also upon doping where magnetic long range order (LRO) sets in. For Ce$_{1-x}$La$_{x}$Ru$_{2}$Si$_{2}$, a QCP is reached at $x_c$=0.075 that separates a paramagnetic ground state ($x$ $<$ $x_c$) with dynamical short range correlations \cite{LPR,Kado0} from an antiferromagnetic ground state ($x$ $>$ $x_c$) with the incommensurate propagation vector $\bf{k_1}$=(0.31, 0, 0)  and the magnetic moments along the tetragonal $c$-axis \cite{Quezel}.

In the present paper, we review recent works performed on this system and give complimentary results obtained on several concentrations $x$=0, 0.065 and 0.075 concerning both the statics and the dynamics in order to tighten the conclusions previously obtained. Data analysis is discussed in the viewpoint of basic properties of the dynamical spin susceptibility and with a minimum reference to a specific lineshape or a specific theory. The scaling approach that was highly developped in recent years is also discussed.

\section{SOFTENING OF THE SPIN FLUCTUATIONS}

The important contribution of INS in magnetism stems from the fact that the scattered neutron intensity, for a given wavevector $\mathbf{q}$ and a given enery $\omega$, is proportional to the scattering function $S(\mathbf{q},\omega)$, from which the imaginary part of the dynamical spin susceptibilty $\chi"(\mathbf{q},\omega)$ is easily deduced using the fluctuation-dissipation theorem :
\begin{equation}
S(\textbf{q},\omega)=\frac{1}{\pi}\frac{1}{1-e^{-\omega/T}}\chi"(\textbf{q},\omega).
\end{equation}
While the main features of the excitation spectrum of the paramagnetic phase of  Ce$_{1-x}$La$_{x}$Ru$_{2}$Si$_{2}$ are known for two decades \cite{LPR}, recent progresses in INS intrumentation (with consequences in data collection rate) allow to map details of the spin dynamics for $x$=0 \cite{Kado0} and $x$=0.075 \cite{KnafoIRIS}.
The softening of the spin fluctuations between $x$=0 and $x$=0.075 was reported a decade ago \cite{RaymondLT} as well as on applying pressure starting from $x$=0.13 \cite{RaymondP}. The microscopic parameters extracted from the dynamical spin susceptibility were used as an input to describe \cite{Kambe} the bulk properties of these compounds with the spin fluctuations theory of Moriya \cite{Moriya}. The imaginary part of the dynamical spin susceptibility, $\chi"(\mathbf{q},\omega)$, is shown in Fig.1 for $x$=0 at 180 mK and $x$=0.075 at 40 mK and 2.3 K, for the momentum transfer $\mathbf{Q}$=(0.69, 1, 0) corresponding to the wavevector $\bf{k_{1}}$ where LRO sets in for $x \textgreater x_c$. The relaxation behaviour of $\chi"(\mathbf{q},\omega)$ for these compounds is well described by a Lorentzian lineshape \cite{KnafoPRB}. No absolute calibration of the intensity is made and the data are normalized to each other at high energy, with the hand-waving argument that the spin dynamics at high energy is, at first approximation, independent of $T$ and $x$. For $x$=0, the maximum ($\omega_{max}$ in Fig.1) of $\chi"(\mathbf{q},\omega)$ is well observed in the experimental window limited by the instrumental resolution (here to $\omega \ge$ 0.25 meV). On the other hand, for the critical concentration $x_{c}$=0.075, only the fall off of $\chi"(\mathbf{q},\omega)$ is measurable (data with better resolution are shown in Ref.\cite{KnafoIRIS} but do not allow to reach $\omega_{max}$ either). This experimental difficulty is a general aspect of the measurement of a critical dynamical spin susceptibility at low temperature by INS (on the contrary NMR gives access to the initial slope for $\omega$ $\rightarrow$ 0). Comparative data similar to the ones shown in Fig.1 are also available for CeCu$_{6-x}$Au$_{x}$ for $x$=0 and the critical concentration $x_{c}$=0.1 (see inset of Fig.4 of Ref.\cite{Stockert}). Study of the spin dynamics of Ce(Ru$_{1-x}$Fe$_{x}$)$_{2}$Ge$_{2}$ (polycrystalline samples) for several $x$=0.65, 0.76, 0.87 around the QCP also put in perspective the interest of such a comparison \cite{Montfrooij2}.

\begin{figure}
\centering\includegraphics[width=6cm]{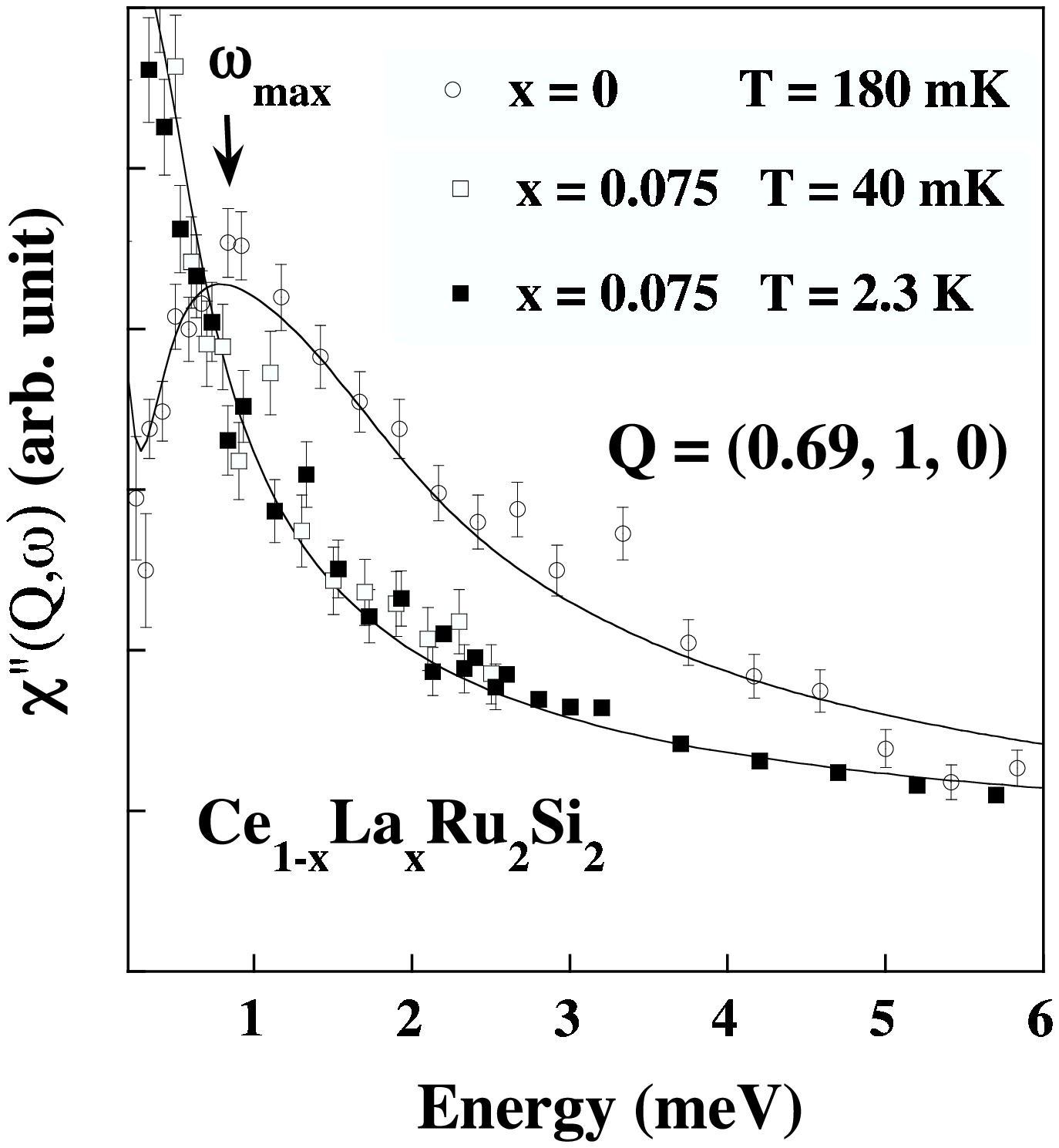}
\caption{Imaginary part of the dynamical spin susceptibility obtained from INS spectra measured at $\bf{k_1}$ at $T$=180 mK for CeRu$_{2}$Si$_{2}$ and $T$=40 mK, 2.3 K for Ce$_{0.925}$La$_{0.0725}$Ru$_{2}$Si$_{2}$.}
\end{figure}

On general ground, $\chi"(\mathbf{q},\omega)$ must go to zero for large $\omega$ and is also an odd function of $\omega$. As a consequence, 
there will always be a maximum of $\chi"(\mathbf{q},\omega)$ at a given value $\omega_{max}$ and this whatever the shape of $\chi"(\mathbf{q},\omega)$
(For a Lorentzian shape, $\omega_{max}=\Gamma$ where $\Gamma$ is the relaxation rate).  Such a peak that does not correspond to a pole of the susceptibility is sometime referred to the paramagnon peak \cite{paramagnon}. The energy $\omega_{max}$ characterizes the crossover between low and high frequency dynamics for gapless systems  \cite{Sachdev}. Its determination is thus somehow not unique ; it leads to several possible data analysis \cite{RaymondLT}. However, it  provides an energy scale for the system at a given reduced wavevector $\mathbf{q}$ and temperature $T$. In the general case, we can write $\omega_{max}(\textbf{q},T)=\omega_{max}(\textbf{q},0)$+$a_{\textbf{q}}$$T^{\beta_{\textbf{q}}}$.
For a given $\mathbf{k}$, the case $\omega_{max}(\textbf{k},0) \rightarrow$0, $\beta_{\mathbf{k}}$=1 corresponds to a so-called scaling in $\omega/T$ of the dynamical spin susceptibility. Such a case is often referred to be characteristic of quantum critical scenario, where there is no other energy scale than the temperature itself at the QCP \cite{Votja,Sachdev}.
A detailed experimental illustration of such a behavior is provided by INS studies on the compound CeCu$_{5.9}$Au$_{0.1}$ \cite{Schroder}. The observation of $\omega/T$ scaling below about 5 K in this compound is however suprising since it is not expected for the QCP of a three dimensional itinerant antiferromagnet that lies above the upper critical dimension. This experimental result led to the developement of a new scenario of quantum criticality, the so-called local scenario \cite{Coleman,Si}, where the local term stems from the experimental fact that the temperature dependence of the  $\mathbf{q}$ dependent susceptibility $\chi'(\textbf{q},T)$ does not depend on $\bf{q}$ (and also $\beta_{\textbf{q}}$ is independent of $\bf{q}$). This is in contrast with itinerant magnetism, for which it is well known that the bulk susceptibility and the staggered susceptibility have different temperature variations at low temperature \cite{Moriya,paramagnon}. The expected relaxational behaviour for a three dimensional itinerant quantum critical antiferromagnet was measured by Kadowaki et al. \cite{KadowakiScaling} in Ce(Ru$_{1-x}$Rh$_{x}$)$_{2}$Si$_{2}$ ($x$=0, 0.03) at the wavevector of instability $\bf{k_3}$=(0, 0, 0.35), where the expected value $\beta_{\mathbf{k_{3}}}$=3/2 is  found below approximately 15 K . A possible explanation for the distinct behaviour at the QCP of CeRu$_{2}$Si$_{2}$ and CeCu$_{6}$ is the different values of the Kondo temperature, being 5-10 times less in CeCu$_{6}$ (See e.g. Table 1 of Ref.\cite{KnafoPRB}).  This asks for very low temperature measurement for CeCu$_{5.9}$Au$_{0.1}$ in order to compare the same temperature regime than in CeRu$_{2}$Si$_{2}$-based systems \cite{Kambe}. Higher temperature relaxation was measured in Ce$_{1-x}$La$_{x}$Ru$_{2}$Si$_{2}$ at $x_c$ \cite{KnafoPRB} in a temperature range (2.5 - 80 K), where $\Gamma(\bf{q})$ $\propto$ $T^{\beta_{\bf{q}}}$. A value $\beta_{\mathbf{k_1}}$=0.8 $\pm$ 0.05 was found for $\bf{k_1}$ while a lower value of 0.6 $\pm$ 0.2 was found for a wavevector away from the critical value, where  Kondo-like fluctuations are expected. In stark contrast, studies on the related system Ce(Ru$_{1-x}$Fe$_{x}$)$_{2}$Ge$_{2}$ found $\beta_{\mathbf{q}}$=1 for every $\bf{q}$ with a behaviour that is neither explained by the local scenario nor by the itinerant scenario \cite{Montfrooij1}. 

Intersite correlations are at the center of magnetic phase transitions (as far as LRO is concerned as opposed to e.g. spin-glass ground state). The discussion of the characteristic  lenght scale (the inverse correlation length, $\kappa$) follows similarly to the discussion of the characteristic energy scale. 
Both quantities are in principle related by the dynamical exponent $z$ with $\Gamma \propto \kappa^z$. While $z$ is important for the dynamics of a classical phase transition \cite{Halperin},
 its importance for a quantum phase transition is already at the level of statics, since it increases the dimension of the system from $d$ to $d_{eff}$.
$\Gamma$ as a function of $\kappa$ is shown in Fig.2 for Ce$_{0.925}$La$_{0.075}$Ru$_{2}$Si$_{2}$, temperature being an implicit parameter. By forcing the condition $\Gamma$=0 for $\kappa$=0, we obtained the value $z$=2 (the result of the fit being $z$=2.1$\pm$ 0.2), as expected for an itinerant antiferromagnet. In such a plot, $\Gamma$ is obtained by the fit of an energy spectrum as in Fig.1 for several temperatures, while $\kappa$ is obtained by fitting consistently many $\bf{q}$ scans performed at constant energy (as the one shown in Fig.6 of Ref.\cite{RaymondLT}). While the exponent $z$ is at the heart of quantum criticality, we are paradoxically not aware of any plot as the one shown in Fig.2  for other HF systems.
\begin{figure}
\centering\includegraphics[width=6cm]{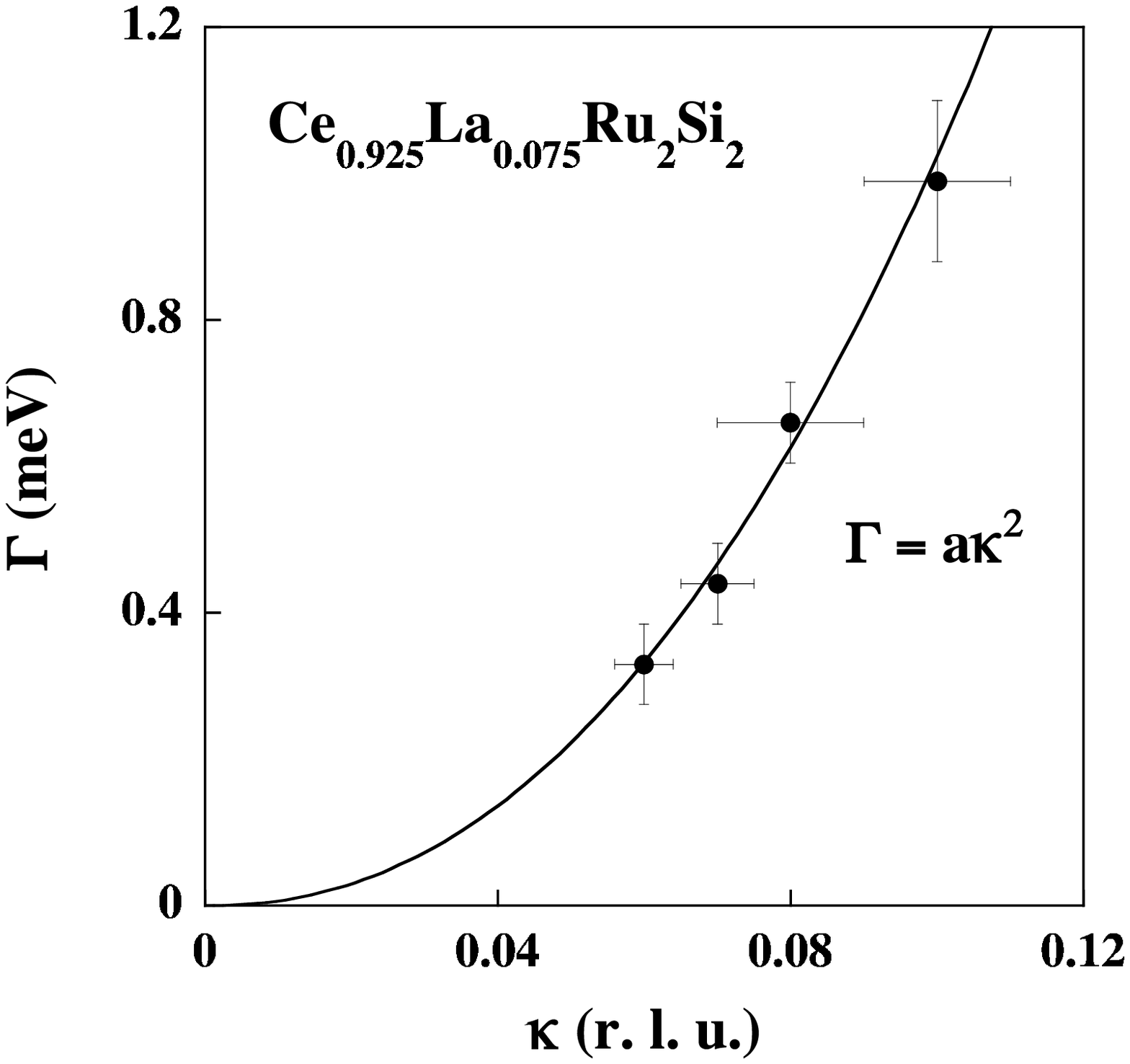}
\caption{Relaxation rate $\Gamma$ as a function of the inverse correlation length $\kappa$ of Ce$_{0.925}$La$_{0.0725}$Ru$_{2}$Si$_{2}$. The line is a fit to $\Gamma$=a$\kappa^2$.}
\end{figure}

In the Ce$_{1-x}$La$_{x}$Ru$_{2}$Si$_{2}$ system, the softening at $x_{c}$ is not complete since the relaxation rate is finite for $T \rightarrow$ 0 and $\bf{q}=\bf{k_1}$ \cite{RaymondLT,KnafoPRB}. This is seen in Fig.1 where the data obtained at $x_{c}$ for 40 mK and 2.3 K are almost identical.  For this reason and also because a tiny static ordered moment appears at $x=x_c$ (see next section), the present system does not provide, at first sight, a "textbook" example of quantum criticality.
Possible explanations for such deviation are that the transition is first order or that other degrees of freedom must be taken into account. The former possibility includes scenarii based on the rich physics of local defects in quantum critical systems \cite{Millis2}. The latter possibility occurs for example in the "textbook" quantum critical compound LiHoF$_{4}$, where the expected mode softening does not occur and is "masked" by hyperfine coupling \cite{Ronnow}. However a study of the behavior of the spin dynamics at high temperature enough will provide relevant information on the so-called quantum critical region, whatever the exact position in $x$ \cite{Votja}.  This is due to the presence, in the $(T,x)$ phase diagram, of two crossover temperatures $T_{I}$ and $T_{II}$ between the quantum disordered regime ($T \textless T_{I}$) and the  quantum critical regime ($T \textgreater T_{II}$) \cite{Sachdev,Millis}.
For heavy fermion systems, the $T_{I}$ line corresponds to the coherence temperature below which a Fermi liquid ground state is formed. Above $T_{I}$, thermal effects arise as perturbations while the regime above $T_{II}$ is completly controlled by the temperature. For a three dimensional itinerant antiferromagnet, the crossover temperatures follow $T_I \propto (x_{c}-x)$ and $T_{II} \propto (x_{c}-x)^{2/3}$. These crossover temperatures can be determined by measuring the temperature dependence of different macrocopic quantities like the specific heat, the resistivity and the susceptibility or of microcopic quantities like the dynamical spin susceptibility and by separating, with choosen criterions, the temperature independent, weakly-temperature dependent and strongly temperature dependent regimes.
Measurements at $x_c$ (where $T_{I}$=0 and $T_{II}$=0) down to $T$=0 K will have similar characteristic exponents than measurements performed for $x \textgreater x_c$ and $T \textgreater T_{II}$. 
For this reason, analysis of the spin fluctuations of CeRu$_{2}$Si$_{2}$ system is sound in the context of quantum criticality. This is emphasized by the fact that the theory of Moriya allows to describe consistently the spin dynamics and the thermodynamic properties of this system \cite{Kambe} and by the recent observation of the characteristic relaxation rate in $T^{3/2}$ for several $x$ near the QCP\cite{KadowakiScaling}. 

\section{FROM TINY TO LARGE MOMENTS}

A small moment ordered phase is reported  for the critical concentration $x_{c}$=0.075 \cite{RaymondLT} with the same ordering vector than the LRO phase and with $m_{0}$=0.02 $\mu_{B}$ , $T_{N}$= 1.8 K and a finite correlation length of about 200 $\AA$. In comparison with $x$=0.08, the N\'eel temperature is similar but the moment is one order of magnitude smaller. For $x$=0, $\mu$SR reports a static moment of 10$^{-3}$ $\mu_{B}$ below 2 K \cite{Amato}. Thus, tiny ordered moments are detected below $x_c$ with however a clear finite correlation length. To further study this phase, a composition $x$=0.065 slightly below $x_{c}$ was investigated.  The development of a small moment starts around the same temperature 4-5 K than the sample with $x$=0.075. This is shown in Fig.3. The left panel shows the elastic peak at 2.7 K as compared to the background at 12 K  and the right panel shows the temperature dependence of the maximum intensity. The peak measured here, although static, has a finite correlation length of 74 $\AA$ (17 Ce-Ce interspacing) at 2.7 K. The signal is proportional to the volume of the sample (as checked by masking several parts) excluding thus macroscopic millimetric heterogeneities. 
\begin{figure}
\centering\includegraphics[width=8cm]{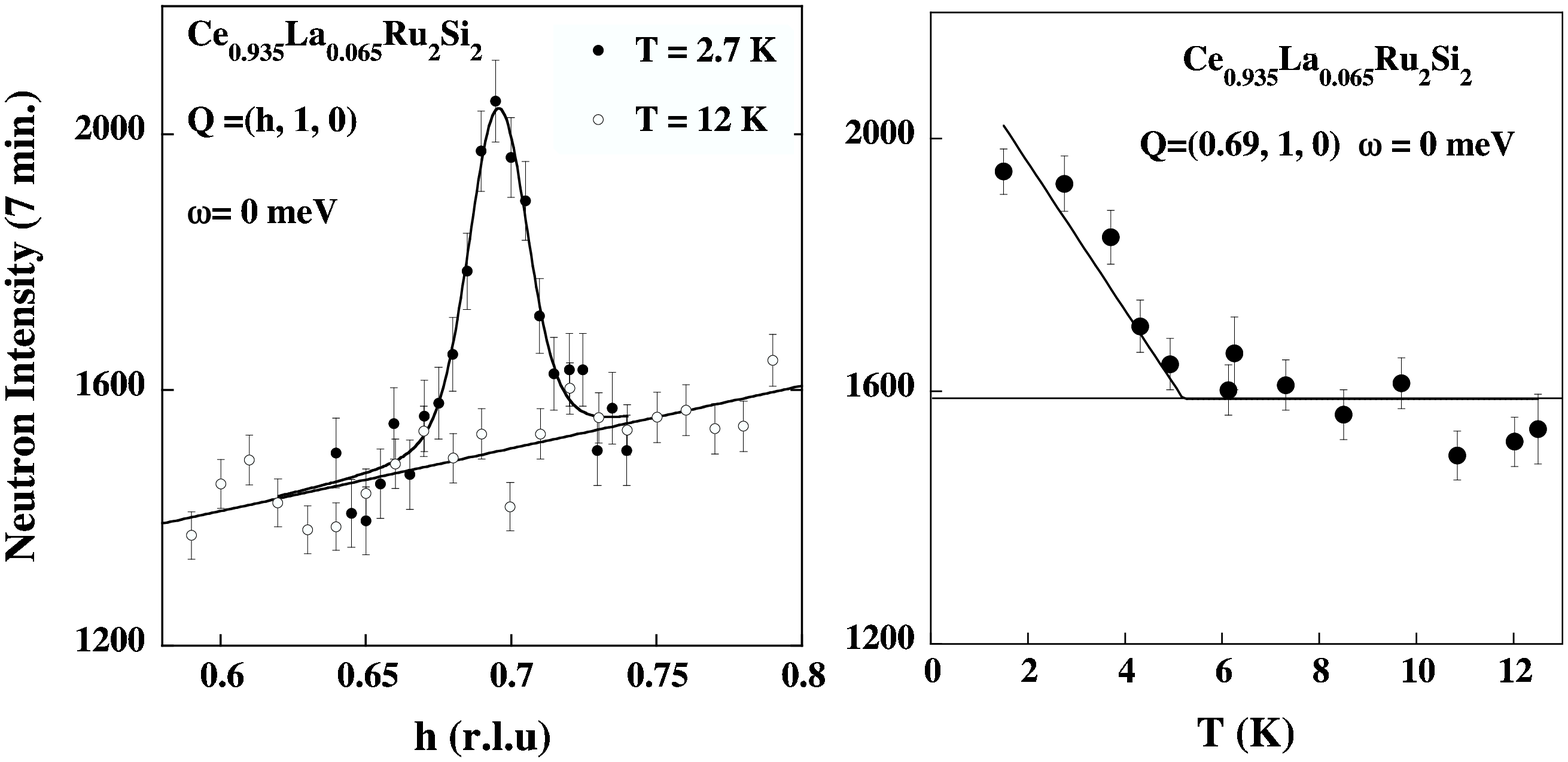}
\caption{Left : Constant energy scans performed at $\omega$=0 meV for $T$=2.7 K and 12 K for Ce$_{0.935}$La$_{0.065}$Ru$_{2}$Si$_{2}$. Right : Temperature dependence of the elastic intensity at $\bf{Q}$=(0.69, 1, 0).}
\end{figure}
Experiments at lower temperature are needed to further investigate the growth of this moment on decreasing temperature in order to determine $T_{N}$ and $m_{0}$, as was done for $x$=0.075. The magnitude of the magnetic moment at 2.7 K is of about the same as in the compound at $x_c$. The intrinsic origin of small moment antiferromagnetism (SMAF) is open (See Ref. \cite{Flouquet2}). Difficult experimental investigations are required. The difficult point is the extreme sensitivity on the pressure and the uniaxial stress close to the QCP. Near defects as dislocations or stacking faults, pressure gradients of kilobar can occur and thus may smear out the real QCP. An illustrative example is that of URu$_{2}$Si$_{2}$, where long standing debates have been made on the intrinsic \cite{Bourdarot} or extrinsic nature of SMAF \cite{Matsuda,Amato2} at low pressure.

For $x > x_c$, LRO develops \cite{Quezel}. Spin fluctuations theory that describes fairly well the spin dynamics of CeRu$_{2}$Si$_{2}$-based alloys\cite{Kado0,Kambe,KadowakiScaling} is less sucessfull  to describe the static properties, in particular the variations of the $T$=0 K ordered magnetic moment, $m_{0}$, and of the N\'eel temperature with $x$, that are expected to follow $m_0 \propto (x-x_c)^{1/2}$ and respectively $T_N \propto (x-x_c)^{2/3}$ (See data in Ref.\cite{RaymondLT}). As a consequence, the following relation is expected : $m_0 \propto T_N^{3/4}$. Figure 4 shows $m_{0}$ as a function of $T_{N}$. The full circles correspond to different $x$ (0.08, 0.1, 0.13, 0.20) and the empty circles correspond to data obtained by applying hydrostatic pressure on the  $x$=0.2 sample \cite{Regnault}. Large uncertainties in the N\'eel temperature are due to rounding of the transitions \cite{Quezel}. Clearly the spin fluctuations prediction, $m_0 \propto T_N^{3/4}$, is not followed.
One interpretation is that the feedback effect of the Kondo screening on the magnetic ordering is not well taken into account (In this theory, the Kondo effect is phenomenologically taken at the first step only as a relaxation mechanism \cite{Moriya}.).
A simple analysis of the reduction of an ordered magnetic moment in presence of the Kondo effect was put forward in a study on CePd$_{2}$Si$_{2}$ \cite{Kernavanois} leading to the following relation :
\begin{equation}
m_{0}=m\sqrt{1-tanh(\Delta/(2T_{N}))}.
\end{equation}
\begin{figure}
\centering\includegraphics[width=6cm]{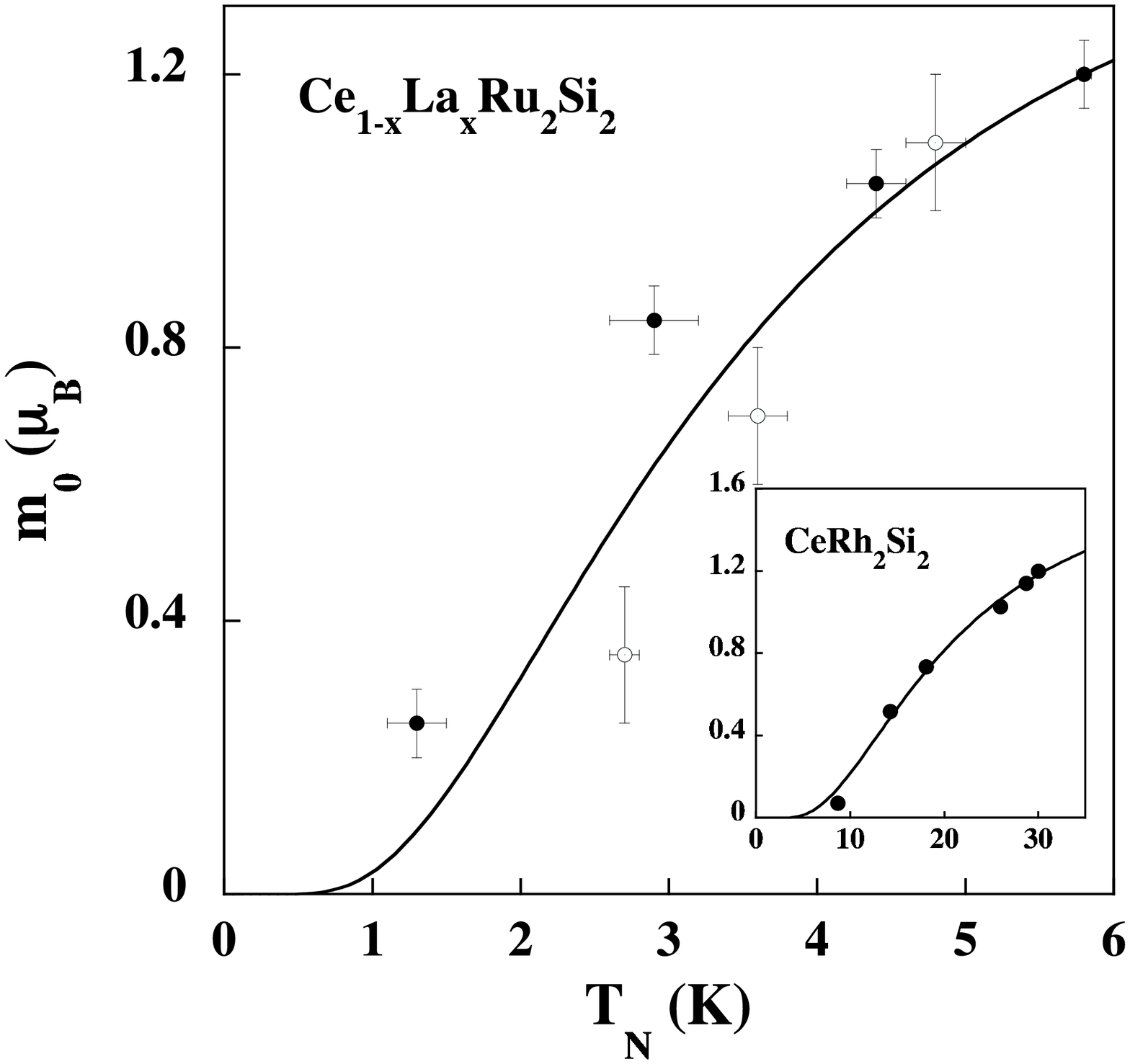}
\caption{$T$=0 K ordered magnetic moment, $m_{0}$, as a function of the N\'eel temperature for Ce$_{1-x}$La$_{x}$Ru$_{2}$Si$_{2}$. Full circles are data for $x$=0.08, 0.1, 0.13, 0.20 and empty circles are data for $x$=0.2 under pressure. The Line is a fit as explained in the text. The inset shows the same curve for CeRh$_{2}$Si$_{2}$.}
\end{figure}
Such a formula is aiming to reproduce the fact that the Kondo effect realizes a singlet by splitting a magnetic Kramers doublet (of paramagnetic moment $m$) by an amount $\Delta \approx k_{B}T_{K}$. The line in Fig 4 corresponds to a fit to (1) and gives $m$=1.6 (0.3) $\mu_{B}$ and $\Delta$= 9 (2) K.
Given the crude model, such values correspond fairly well to the crystal field ground state magnetic moment of the CeRu$_{2}$Si$_{2}$ (2 $\mu_{B}$) and the average Kondo temperature of 15 K over the serie. In such a picture, a large Kondo temperature will lead to the possibility of having small moments with large N\'eel temperatures. However the SMAF found for $x$=0 and $x$=0.075 cannot clearly fit in this picture and needs another mechanism. On the contrary, the change of behavior from large to small moment pointed forward for CeRh$_{2}$Si$_{2}$ under pressure\cite{Kawarazaki} can be, to this respect, understood from (1) given the large Kondo temperature of this compound of about 50 K.
This is shown in the inset of Fig.4 where the data of Ref\cite{Kawarazaki} are fitted by (1) giving $m$=1.8 (0.1) $\mu_{B}$ and $\Delta$= 55 (3) K. Here again the pertinence of the obtained parameters is suprising for a such rough phenomenological model.

\section{BEHAVIOUR AWAY FROM QCP}

In Ref.\cite{KnafoPRB}, it was found that the dynamical spin susceptibility of Ce$_{0.925}$La$_{0.075}$Ru$_{2}$Si$_{2}$ follows, for the wavevector of instability $\bf{k_1}$ and for 2.5 K  $\textless$ $T$ $\textless$  80 K, the scaling law :
\begin{equation}
T\chi"({\mathbf{k_1}},\omega)=C_{1}f(\omega/T^{0.8}).
\end{equation}
This scaling does not hold away from $\bf{k_1}$. For the momentum transfer $\bf{Q}$=(0.44, 1, 0) (that will be referred as wavevector $\bf{q_0}$ in the following), it is found that $\Gamma(\textbf{q}_0,T) \propto T^{0.6 \pm 0.2}$. Also the characteristic energy scale at this wavevector, $\Gamma(\textbf{q}_0,0)$, is about  7 times higher than the one at $\bf{k_1}$, so that a sound study will need to achieve temperature for which higher energy levels may be involved (typically crystal field like excitation) and will interfere with the ground state behavior. As a consequence scaling behavior cannot be put in perspective for this $\bf{q}$ (See Ref.\cite{KnafoPRB} for further details).
\begin{figure}
\centering\includegraphics[width=6cm]{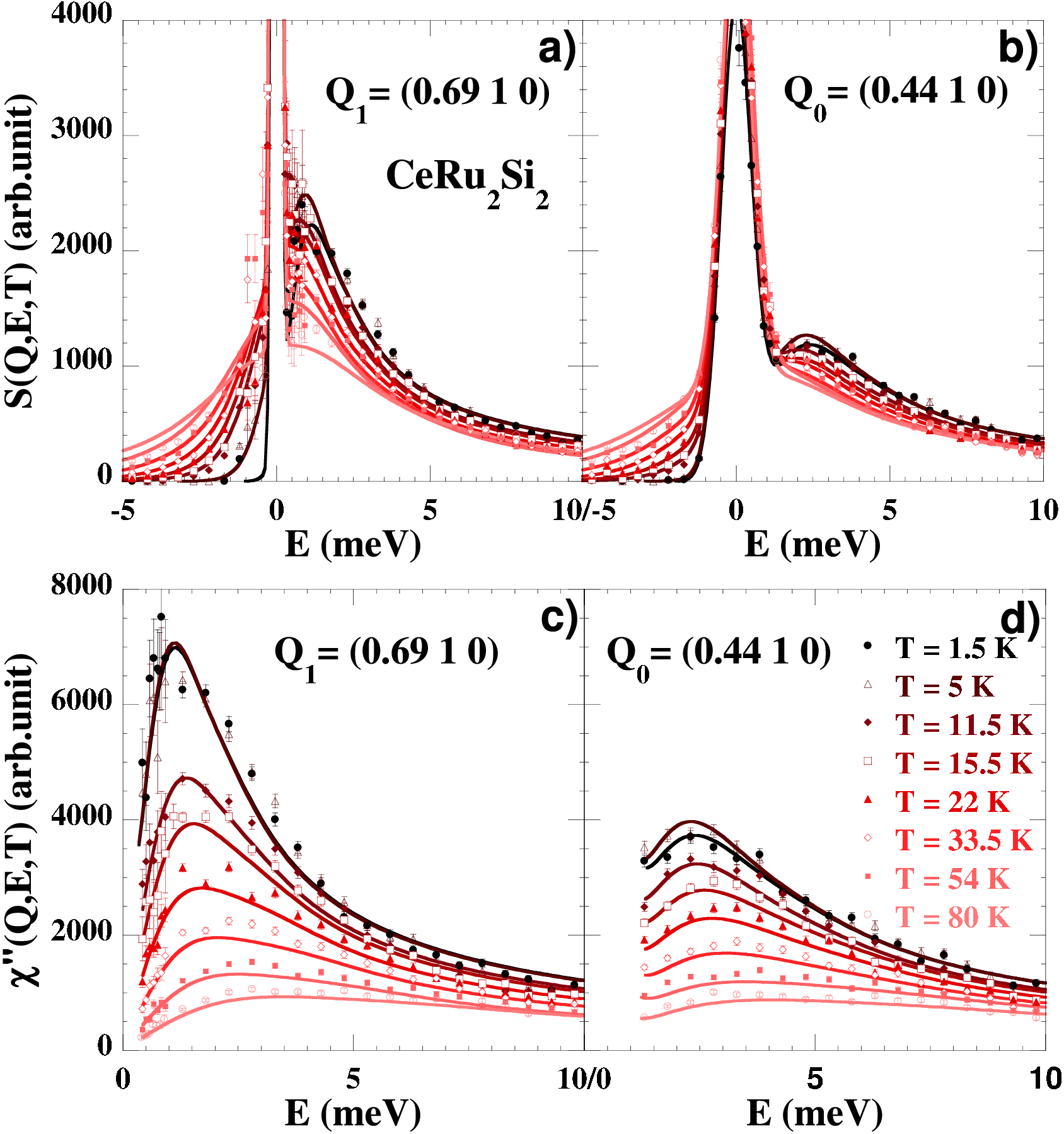}
\caption{INS spectra obtained of CeRu$_{2}$Si$_{2}$ for $\bf{k_1}$ (a) and $\bf{q_0}$ (b) and corresponding imaginary part of the dynamical spin susceptibility (c) and (d) for 1.5 K $\textless T \textless$ 80 K.}
\end{figure}
An important question to address is to which respect this behavior is peculiar to the QCP. Thus, new experiments on CeRu$_{2}$Si$_{2}$ were undertaken \cite{Knafothese} in the spirit of the work of Ref.\cite{KnafoPRB}, that is by exploring the dynamical spin susceptibility in wide $\omega$ and $T$ ranges. Bare data of $S(\bf{q},\omega)$ and $\chi"(\bf{q},\omega)$ are shown in Fig.5 for the critical wavevector $\bf{k_1}$ and for the wavevector $\bf{q_0}$ for 2.5 K $\textless T \textless$ 80 K. This is to be compared to Fig.2 of Ref.\cite{KnafoPRB}. At low temperature, the signal is stronger at the vector $\bf{k_1}$ and this difference in intensity between the two vectors progressively decreases and vanishes at $T_{corr} \approx $ 80 K, when correlations disappear. The data analysis is the same as in Ref.\cite{KnafoPRB}. The corresponding parameters extracted from a fit to a Lorentzian are shown in Fig. 6 on the left panel for $\bf{k_{1}}$ and on the right panel for $\bf{q_{0}}$. For both $\bf{q}$ values, strong temperature dependence of $\chi'(\textbf{q},T)$ ($\chi'(\textbf{q},T)$ is obtained by Kramers Kronig relation\cite{KnafoPRB}) and $\Gamma(\textbf{q},T)$ is found with power law behaviour for $T$ $\textgreater$ $T_{\textbf{q}}$ :
\begin{equation}
\chi'(\textbf{q},T)= C_{\mathbf{q}}/T^{\alpha_{\mathbf{q}}}
\end{equation}
\begin{equation}
\Gamma(\textbf{q},T)=a_{\bf{q}}T^{\beta_{\textbf{q}}}
\end{equation}
with $T_{\mathbf{k_1}}$=9 $\pm$ 1 K , $\alpha_{\mathbf{k_1}}$=1 $\pm$ 0.1, $\beta_{\mathbf{k_1}}$=0.7 $\pm$ 0.2 and $T_{\mathbf{q_0}}$= 22 $\pm$ 2 K, $\alpha_{\mathbf{q_o}}$=1 $\pm$ 0.1, $\beta_{\mathbf{q_0}}$=0.6 $\pm$ 0.2.
\begin{figure}
\centering\includegraphics[width=8cm]{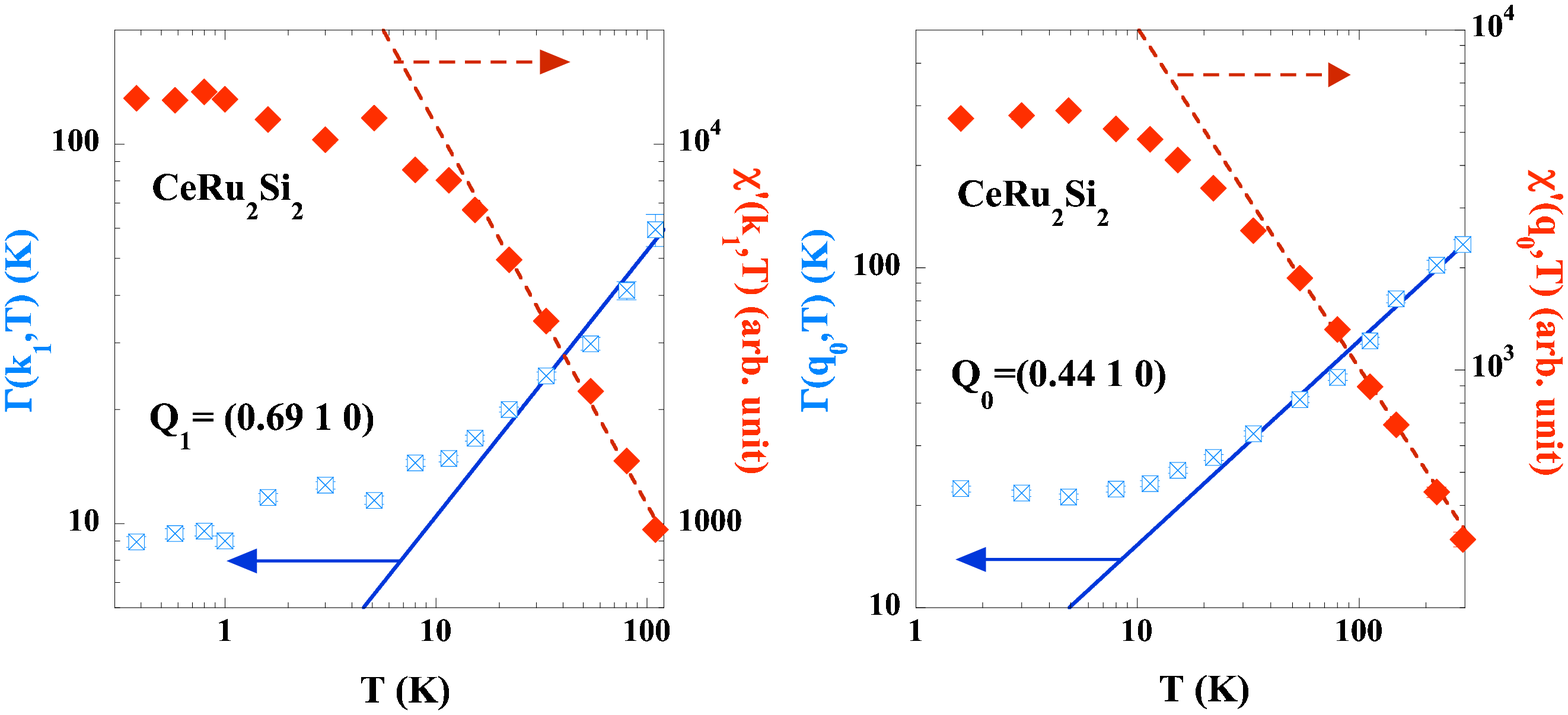}
\caption{Temperature variation of the relaxation rate and of the $\textbf{q}$-dependent static susceptibility for $\bf{k_1}$ (left) and $\bf{q_0}$ (right) for CeRu$_{2}$Si$_{2}$. Lines correspond to adjustments with (4) and (5).}
\end{figure}

For the same reasons as the one discussed above for the wavevector $\bf{q_{0}}$ for $x=x_c$, scaling cannot be reasonably discussed from these data obtained for $x$=0 at both $\bf{k_1}$ and $\bf{q_0}$. For each $\bf{q}$, sublinear $T$-dependence of $\Gamma(\textbf{q},T)$ seems to be characteristic of the system Ce$_{1-x}$La$_{x}$Ru$_{2}$Si$_{2}$. For $x$=0, the $\bf{q}$ dependence of $\beta_{\mathbf{q}}$ is not as obvious as in the sample at the critical concentration and cannot be discussed here due to the weaker signal and the large error bars for $\beta_{\mathbf{q}}$. The different exponents and characteristic temperatures obtained for different $\bf{q}$ for $x$=0 and $x$=0.075 are summarized in Table 1.
\begin{table}
\caption{Characteristic temperature $T_{\mathbf{q}}$ and
exponents $\alpha_{\bf{q}}$ and $\beta_{\bf{q}}$  of power law dependence of $\chi'(\textbf{q},T)$ and
$\Gamma(\textbf{q},T)$ for the compounds Ce$_{0.925}$La$_{0.075}$Ru$_{2}$Si$_{2}$ and CeRu$_{2}$Si$_{2}$ at
wavevector $\bf{k}_{1}$, $\bf{q}_{0}$ and $\bf{q}=0$.}
\begin{center}
\item[]\begin{tabular}{|c||c|c|c|c|c|c|}
\cline{2-7} \multicolumn{1}{l|}{} & \multicolumn{3}{c}{Ce$_{0.925}$La$_{0.075}$Ru$_{2}$Si$_{2}$} &  \multicolumn{3}{|c|}{CeRu$_{2}$Si$_{2}$} \\
\multicolumn{1}{l|}{} & $\mathbf{k}_{1}$ &  $\mathbf{q}_{0}$ & $\mathbf{q}=0$ $\,^{*}$ & $\mathbf{k}_{1}$ &  $\mathbf{q}_{0}$ & $\mathbf{q}=0$ $\,^{*}$ \\
\hline
$T_{\mathbf{q}}$ & 2.5 $\pm$ 0.5 K& 18 $\pm$ 1 K & 16 $\pm$ 1 K & 9 $\pm$ 1 K & 22 $\pm$ 2 K & 28 $\pm$ 2 K \\
 $\alpha_{\mathbf{q}}$ & $1\pm0.05$  &  $1\pm0.1$ & 1 &  $1\pm0.1$&  $1\pm0.1$ & 1 \\
 $\beta_{\mathbf{q}}$ &  $0.8\pm0.05$ & $0.6\pm0.2$ & - & $0.7\pm0.2$ & $0.6\pm0.2$ & -\\
\hline
\end{tabular}
\end{center}
$\,^{*}$ : \footnotesize{after $\chi_{bulk}(T)$ from Ref.  \cite{Knafothese}}\\
\label{celarusiTalphabeta}
\end{table}
The systematics indicates that :  for each $x$ and $\bf{q}$, there is a weakly temperature dependent regime below $T_{\bf{q}}$ where $\Gamma(\textbf{q},T)$ $\approx$ $T_{\bf{q}}$ and $\chi'(\textbf{q},T)$ $\approx$ cte. 
Normalization at high temperature with the the bulk susceptibility indicates that $T_{\textbf{q}}\chi'(\textbf{q},T) \approx$ 1 K.emu.mol$^{-1}$ for each $x$ and $\bf{q}$.
Above $T_{\bf{q}}$, laws corresponding to (3) and (4) are found.  In particular, $\chi'(\textbf{q},T) \propto 1/T$ for any $x$ and $\textbf{q}$, this latter law being also found for the bulk susceptibility \cite{KnafoPRB,Knafothese}. On another hand, the exponent $\beta_{\mathbf{q}}$ is distinctly higher for $x_{c}$ and $\bf{k_1}$ than for the other cases with $\beta_{\mathbf{k_1}}$=0.8 $\pm 0.05$. This could lead support to the idea that getting away from the QCP with varying $x$ at $\bf{k_1}$ or with varying $\bf{q}$ at $x_c$ is somehow equivalent.  This idea was also put forward for Ce(Ru$_{1-x}$Fe$_{x}$)$_{2}$Si$_{2}$ \cite{Montfrooij2}. The discussion of the weakly temperature dependent regime ($T$ $<$ $T_{\bf{q}}$) requires excellent statistics to point out the characteritics of the quantum critical regime. As pointed out before, this aim was realized in Ce(Ru$_{1-x}$Rh$_{x}$)$_{2}$Si$_{2}$ ($x$=0, 0.03) \cite{KadowakiScaling}.

\section{SCALING ANALYSIS}

The experimental limitation on the access of the low energy part of  $\chi"(\textbf{q},\omega)$, as discussed in the first part, implies to our opinion that the discussion of the lineshape of the spectra is very difficult, at least at the lowest temperature. Instead of analyzing each spectra as done above, it has been fashionable to study the data collapse as a function of a single variable, being generally $\omega/T$, and to address the question of the lineshape after performing this global analysis. The limit of this latter approach was evidenced in Ref.\cite{KnafoSCALING} in which several possible scalings are shown for the same given data set. We shall further comment the scaling approach here by comparing cases available in the literature. Beyond the cases already quoted, scaling of the dynamical spin suceptibility was also reported for UCu$_{5-x}$Pd$_{x}$\cite{Aronson}, Sc$_{1-x}$U$_{x}$Pd$_{3}$ \cite{Wilson}, Ce(Rh$_{0.8}$Pd$_{0.2}$)Sb \cite{Park} and (Ce$_{0.7}$Th$_{0.3}$)RhSb\cite{So}. Beside cerium and uranium based HF, recent INS studies performed on single crystals of La$_{2}$CuO$_{4}$-based systems also discuss the scaling behavior in $\omega/T$ of the dynamical spin susceptibility \cite{Bao,Chen}.

We are interested here in the general notion of QCP and the possibility of $\omega/T$ scaling irrespective of the different mechanisms at play in the variety of systems referred above. The differences are mainly the dimensionality of the system and of the order parameter, the nature of the ordered ground state (magnetic LRO or spin-glass order) and the $\bf{q}$-dependence of the spin dynamics. Here we will only focus on the energy dependence at the vector of instability or at any vector for the systems without correlations.
An alternate expression, better suited for the following discussion,  of the fluctuation-dissipation theorem (1) is :
\begin{equation}
\chi"(\bf{q},\omega)=\pi[S(\bf{q},\omega)-S(\bf{q},-\omega)].
\end{equation}
The formula (1) is derived from this more general expression (6) by using the detailed balance principle for centrosymetric systems ($\bf{q}$ reversal) :
\begin{equation}
S(\textbf{q},\omega)=e^{\omega/T}S(\textbf{q},-\omega).
\end{equation}
For UCu$_{5-x}$Pd$_{x}$ ($x$=1) \cite{Aronson} and Sc$_{1-x}$U$_{x}$Pd$_{3}$ ($x$=0.35) \cite{Wilson}, it is reported that $S(\bf{q},\omega)$ is independent of temperature. As a consequence, the main temperature comes from the thermal population factor, so that $\omega/T$ scaling is de-facto obtained for $\chi"(\textbf{q},\omega)$. If furthermore 
$S(\textbf{q},\omega) \propto 1/\omega^{\alpha}$, the Kramers Kronig relation implies that $\chi_{bulk} \propto 1/T^{\alpha}$, that could correspond to marginal Fermi liquid since this will be observed in a wide temperature range, accordingly to the behaviour of $S(\textbf{q},\omega)$.
The same "$T$-independent $S(\bf{q},\omega)$" scenario is realized in Ce(Rh$_{0.8}$Pd$_{0.2}$)Sb \cite{Park} and (Ce$_{0.7}$Th$_{0.3}$)RhSb \cite{So}.
Given equation (6), it is quite surprising that the temperature dependences of $S(\bf{q},\omega)$ and $\chi"(\bf{q},\omega)$ are so different. All the more textbook example of fluctuations process indicates that it is rather $\chi"(\bf{q},\omega)$ that does not need to depend on temperature while $S(\bf{q},\omega)$ does \cite{Nozieres}. The profond physical meaning of the collapse of the data in $S(\bf{q},\omega)$ for $\omega \textgreater$ 0 in a $T$-independent way and to which extend this is peculiar to some class of QCP needs to be tighten. 

At opposite, in the cases of  CeCu$_{6}$ and CeRu$_{2}$Si$_{2}$ based compounds, as well as for  the CuO$_{2}$ based systems recently investigated\cite{Bao,Chen}, both $S(\bf{q},\omega)$ and $\chi"(\bf{q},\omega)$ are temperature dependent (See Fig.2 of \cite{KnafoPRB} and Fig.5 of the present paper). For CeRu$_{2}$Si$_{2}$-based compounds, if only data with $\omega \textgreater$ 5 meV are considered, $S(\bf{q},\omega)$ is independent of $T$ (See Fig.5) as for the cases treated in the previous paragraph. However substantial $T$ dependence occurs at low energy. 
In the case of CeCu$_{6}$, scaling is even made either on $S(\bf{q},\omega)$ \cite{Schroder} or $\chi"(\bf{q},\omega)$ \cite{schroderold}, with the same physical conclusions on the exponents. In these two compounds,  thermal population factor is thus not the main origin of the scaling as in the former systems. 

\section{COMPARISON BETWEEN CeRu$_{2}$Si$_{2}$-BASED SYSTEMS}

\begin{table}
\caption{Different exponents $\beta_{\mathbf{q}}$ of the temperature dependence of the relaxation rate $\Gamma(\textbf{q},T)$ obtained for several CeRu$_{2}$X$_{2}$-based systems (X=Si,Ge) at their respective QCP for the wavevector of instability (first three lines) and for CeRu$_{2}$Si$_{2}$ away from the wavevector of instability (last two lines).}
\begin{center}
\item[]
\begin{tabular}{|c|c|c|c|}
\hline
Compound  & Exponent $\beta_{\mathbf{q}}$  &  $T$-range of analysis & Reference \\
\hline
Ce(Ru$_{0.97}$Rh$_{0.03}$)$_{2}$Si$_{2}$  & 1.5 & 1.5 K $\textless T \textless$ 8 K  & Ref.\cite{KadowakiScaling} \\
Ce(Ru$_{0.24}$Fe$_{0.76}$)$_{2}$Ge$_{2}$ &  1 & 1.9 K $\textless T \textless$ 200 K &  Ref.$^{*}$\cite{Montfrooij1}  \\
 Ce$_{0.925}$La$_{0.075}$Ru$_{2}$Si$_{2}$ &  0.8 &  5 K $\textless T \textless$ 80 K & Ref.\cite{KnafoPRB}  \\
\hline
 CeRu$_{2}$Si$_{2}$ & 0.5 & 10 K $\textless T \textless$ 200 K &  Ref.\cite{Loidl} \\
 CeRu$_{2}$Si$_{2}$ & 0.6 & 22 K $\textless T \textless $ 80 K &  Ref.\cite{Knafothese}\\
\hline
\end{tabular}
\end{center}
$\,^{*}$ : \footnotesize{With non Lorentzian lineshape data analysis}
\end{table}

While doping Ce with La or Ru with Rh, starting from CeRu$_{2}$Si$_{2}$, or doping Ru with Fe, starting from CeRu$_{2}$Ge$_{2}$, (the relation between CeRu$_{2}$Si$_{2}$ and CeRu$_{2}$Ge$_{2}$ is for exemple given in Ref.\cite{Heribert}) is expected to give similar results at first approximation, data reported in the literature are quite different \cite{KnafoPRB,Montfrooij2,KadowakiScaling,Montfrooij1}. One reason is that the temperature ranges considered are different, as well as the data analysis procedures. Also the role of disorder must be taken into account since for a given system like Ce(Ru$_{1-x}$Rh$_{x}$)$_{2}$Si$_{2}$ ,  different behaviours were evidenced between small $x$\cite{Tabata1} (CeRu$_{2}$Si$_{2}$ side) and large $x$\cite{Tabata2} (CeRh$_{2}$Si$_{2}$ side). 
For a clean system, in the paramagnetic phase around the QCP, the relaxation is expected to follow a Fermi-liquid law below $T_I$ : $\Gamma(\textbf{k},T)$=$\Gamma(\textbf{k},0)+a_{\textbf{k}}T^{2}$. Above $T_{II}$, a $T^{3/2}$ law is expected contrarily to the naive expectation of $T$ linear relaxation due to the fact that the system lies above its upper critical dimension (intermediate behavior is expected between $T_I$ and $T_{II}$). At higher temperature, Kondo relaxation is well known to occur with a characteristic $\sqrt(T)$ law \cite{Loidl,Cox}. However the influence of the line $T_K$ is not taken into account in the theories of itinerant QCP. In the local scenario of QCP\cite{Coleman}, it has even been proposed that the reconstruction of the Fermi surface will push $T_{K}$ to zero. All these compounds have rather similar Kondo temperature 15-20 K except for Ce(Ru$_{0.24}$Fe$_{0.76}$)$_{2}$Ge$_{2}$ for which $T_{K}$=5 K\cite{Montfrooij1}. This could explain the linear temperature law found in this compound in the same line than for CeCu$_{6}$-based system, for which $T_{K}$=3 K. The first relaxation regime below $T_{I}$ was never evidenced to our knowledge certainly because already for pure CeRu$_{2}$Si$_{2}$, $T_I \approx$ 1 K. The second regime is well investigated by Kadowaki et al. on Rh-doped samples \cite{KadowakiScaling}. Scaling procedure, that aims to evidence large temperature dependence, smears out this regime.Thus,  one simple explanation for the $T^{0.8}$ or $T$ behavior observed respectively for La-doped sample\cite{KnafoPRB} and Fe-doped sample\cite{Montfrooij1}, is to consider this as a crossover between the $T^{3/2}$ regime and $\sqrt(T)$ regime. Table 2 summerizes the different regimes experimentally evidenced for CeRu$_{2}$Si$_{2}$-based systems. Certainly, the temperature exponent  is higher when data are taken at lower temperature in a small range and it decreases when the temperature window increases in the high temperature side. Beyond the differences in characteristic energy scales $T_{K}$ and in the temperature window of investigation of the dynamical spin susceptibility, the sample quality must be taken into account. To this respect, the residual resistivity is a good criterion. While it is of the order of 5-10 $\mu\Omega$cm for Ce$_{0.925}$La$_{0.0725}$Ru$_{2}$Si$_{2}$\cite{Kambe} and Ce(Ru$_{0.97}$Rh$_{0.03}$)$_{2}$Si$_{2}$\cite{Miyako}, higher values of 50-75 are found for Ce(Ru$_{0.24}$Fe$_{0.76}$)$_{2}$Ge$_{2}$\cite{Montfrooij3}. Such differences could also be at the origin of the different behaviour at the QCP of these compounds. In the limit of strong disorder,  Griffith phases may occur with stronger effect at the QCP as for classical phase transition \cite{Vojta2}. As concern the spin dynamics, strong disorder could lead to distribution of relaxation rates with different possible models characterized by non-Lorentzian lineshape \cite{Nick}.

\section{CONCLUSION}
The study of CeRu$_{2}$Si$_{2}$ based compounds clearly underlines the role of the Kondo effect in the relaxation spectrum at the QCP. This is mainly shown by the spectral weight dominated by Kondo $\bf{q}$-independent fluctuations and by the sublinear relaxation rate obtained at high temperature. In the present paper, importance of the Kondo effect for the statics is also pointed out, given a mechanism for strong reduction of the static moment.  The saturation of the relaxation rate at the QCP and the recurent presence of tiny ordered moment have at present unknown origin. The overall study of Ce$_{1-x}$La$_{x}$Ru$_{2}$Si$_{2}$ stresses on the need to  compare different $x$ to pinpoint the peculiarity of the behavior at the critical concentration. Moreover the publication of bare data as well as  analysis of single spectra is highly desirable. This would allow to get round the main difficulty of the topic that lies in the fact that each group analyses its own data with different procedures, making a global comparison between several samples extremely difficult.

The work of Pr. L\"ohneysen group on CeCu$_{6-x}$Au$_{x}$ strongly stimulated us to perform a continuous experimental program on Ce$_{1-x}$La$_{x}$Ru$_{2}$Si$_{2}$ over two decades.

\section{Acknowledgement}
This work was performed at Institut Laue Langevin, Grenoble on the CEA-CRG spectrometers IN12 and IN22 and at Laboratoire L\'eon Brillouin (UMR CEA-CNRS) on the spectrometers 4F1/4F2.

\end{document}